\pdfoutput=1
\documentclass[12pt,a4paper]{article}

\usepackage{ifthen} 
\newboolean{pdflatex}
\setboolean{pdflatex}{true} 

\newboolean{articletitles}
\setboolean{articletitles}{true} 

\newboolean{uprightparticles}
\setboolean{uprightparticles}{false} 
\usepackage{decays} 

\def\paperauthors{P.~Koppenburg} 
\def\paperasciititle{Flavour Physics beyond the LHC} 
\def\papertitle{Flavour Physics beyond the LHC} 
\def\paperkeywords{{High Energy Physics}} 
\def\papercopyright{\the\year} 
\def\paperlicence{CC BY 4.0 licence}
\def\paperlicenceurl{https://creativecommons.org/licenses/by/4.0/}

\newif\ifEnableSectionTOCLinks
\EnableSectionTOCLinksfalse 

\usepackage[top=1in, bottom=1.25in, left=1in, right=1in]{geometry}

\usepackage{microtype}
\usepackage{lineno}  
\usepackage{xspace} 
\usepackage{caption} 

\usepackage{graphicx}  
\usepackage{color}
\usepackage{colortbl}
\graphicspath{{./figs/}} 

\usepackage{amsmath} 
\usepackage{amssymb}
\usepackage{amsfonts}
\usepackage{upgreek} 

\newcommand*\patchAmsMathEnvironmentForLineno[1]{%
\expandafter\let\csname old#1\expandafter\endcsname\csname #1\endcsname
\expandafter\let\csname oldend#1\expandafter\endcsname\csname
end#1\endcsname
 \renewenvironment{#1}%
   {\linenomath\csname old#1\endcsname}%
   {\csname oldend#1\endcsname\endlinenomath}%
}
\newcommand*\patchBothAmsMathEnvironmentsForLineno[1]{%
  \patchAmsMathEnvironmentForLineno{#1}%
  \patchAmsMathEnvironmentForLineno{#1*}%
}
\AtBeginDocument{%
\patchBothAmsMathEnvironmentsForLineno{equation}%
\patchBothAmsMathEnvironmentsForLineno{align}%
\patchBothAmsMathEnvironmentsForLineno{flalign}%
\patchBothAmsMathEnvironmentsForLineno{alignat}%
\patchBothAmsMathEnvironmentsForLineno{gather}%
\patchBothAmsMathEnvironmentsForLineno{multline}%
\patchBothAmsMathEnvironmentsForLineno{eqnarray}%
}

\usepackage[pdftex,
            pdfauthor={\paperauthors},
            pdftitle={\paperasciititle},
            pdfkeywords={\paperkeywords}]{hyperref}
\usepackage{hyperxmp}
\hypersetup{
    pdfcopyright={Copyright (C) \papercopyright},
    pdflicenseurl={\paperlicenceurl}
}

\usepackage[colorinlistoftodos,textsize=scriptsize]{todonotes}

\usepackage[bottom,flushmargin,hang,multiple]{footmisc}

\usepackage[all]{hypcap} 

\usepackage{xspace} 
\usepackage{upgreek}

\def\MagUp {\mbox{\em Mag\kern -0.05em Up}\xspace}

\ifthenelse{\boolean{uprightparticles}}%
{
 
 \def\Pgamma      {\ensuremath{\upgamma}\xspace}

 \def\Pmu         {\ensuremath{\upmu}\xspace}                 
 \def\Pnu         {\ensuremath{\upnu}\xspace}                 
                  
 \def\Ppi         {\ensuremath{\uppi}\xspace}

 \def\Ptau        {\ensuremath{\uptau}\xspace}                 
                  
 \def\Pphi        {\ensuremath{\upphi}\xspace}

 \def\Ppsi        {\ensuremath{\uppsi}\xspace}

 \def\PDelta      {\ensuremath{\Delta}\xspace}                 
 \def\PXi         {\ensuremath{\Xi}\xspace}                 
 \def\PLambda     {\ensuremath{\Lambda}\xspace}                 
 \def\PSigma      {\ensuremath{\Sigma}\xspace}                 
 \def\POmega      {\ensuremath{\Omega}\xspace}                 
 \def\PUpsilon    {\ensuremath{\Upsilon}\xspace}
 \let\oldPi\Pi
 \def\PPi         {\ensuremath{\oldPi}\xspace}

 \def\PB      {\ensuremath{\mathrm{B}}\xspace}                 
 \def\PD      {\ensuremath{\mathrm{D}}\xspace}                 
 \def\PJ      {\ensuremath{\mathrm{J}}\xspace}                 
 \def\PK      {\ensuremath{\mathrm{K}}\xspace}                 
 \def\PW      {\ensuremath{\mathrm{W}}\xspace}                 
 \def\PZ      {\ensuremath{\mathrm{Z}}\xspace}                 
 \def\Pb      {\ensuremath{\mathrm{b}}\xspace}                 
 \def\Pc      {\ensuremath{\mathrm{c}}\xspace}                 
 \def\Pd      {\ensuremath{\mathrm{d}}\xspace}                 
 \def\Pe      {\ensuremath{\mathrm{e}}\xspace}                 
 \def\Pp      {\ensuremath{\mathrm{p}}\xspace}                 
 \def\Pq      {\ensuremath{\mathrm{q}}\xspace}                 
 \def\Ps      {\ensuremath{\mathrm{s}}\xspace}                 
 \def\Pt      {\ensuremath{\mathrm{t}}\xspace}                 
 \def\Pu      {\ensuremath{\mathrm{u}}\xspace}                 
 \def\thebaroffset{0.0em}
}
{
 
 \def\Pgamma      {\ensuremath{\gamma}\xspace}

 \def\Pmu         {\ensuremath{\mu}\xspace}                 
 \def\Pnu         {\ensuremath{\nu}\xspace}                 
                  
 \def\Ppi         {\ensuremath{\pi}\xspace}

 \def\Ptau        {\ensuremath{\tau}\xspace}                 
                  
 \def\Pphi        {\ensuremath{\phi}\xspace}

 \def\Ppsi        {\ensuremath{\psi}\xspace}                 
                  
 \mathchardef\PDelta="7101
 \mathchardef\PXi="7104
 \mathchardef\PLambda="7103
 \mathchardef\PSigma="7106
 \mathchardef\POmega="710A
 \mathchardef\PUpsilon="7107
 \mathchardef\PPi="7105
 \def\PB      {\ensuremath{B}\xspace}                 
 \def\PD      {\ensuremath{D}\xspace}                 
 \def\PJ      {\ensuremath{J}\xspace}                 
 \def\PK      {\ensuremath{K}\xspace}                 
 \def\PW      {\ensuremath{W}\xspace}                 
 \def\PZ      {\ensuremath{Z}\xspace}                 
 \def\Pb      {\ensuremath{b}\xspace}                 
 \def\Pc      {\ensuremath{c}\xspace}                 
 \def\Pd      {\ensuremath{d}\xspace}                 
 \def\Pe      {\ensuremath{e}\xspace}                 
 \def\Pp      {\ensuremath{p}\xspace}                 
 \def\Pq      {\ensuremath{q}\xspace}                 
 \def\Ps      {\ensuremath{s}\xspace}                 
 \def\Pt      {\ensuremath{t}\xspace}                 
 \def\Pu      {\ensuremath{u}\xspace}                 
 \def\thebaroffset{0.18em}
}
\newcommand{\offsetoverline}[2][\thebaroffset]{\kern #1\overline{\kern -#1 #2}}%

\makeatletter
\ifcase \@ptsize \relax
  \newcommand{\miniscule}{\@setfontsize\miniscule{4}{5}}
\or
  \newcommand{\miniscule}{\@setfontsize\miniscule{5}{6}}
\or
  \newcommand{\miniscule}{\@setfontsize\miniscule{5}{6}}
\fi
\makeatother

\DeclareRobustCommand{\optbar}[1]{\shortstack{{\miniscule (\rule[.5ex]{1.25em}{.18mm})}
  \\ [-.7ex] $#1$}}

\def\epem       {{\ensuremath{\Pe^+\Pe^-}}\xspace}

\def\muon       {{\ensuremath{\Pmu}}\xspace}
\def\mup        {{\ensuremath{\Pmu^+}}\xspace}
\def\mun        {{\ensuremath{\Pmu^-}}\xspace} 

\def\mumu       {{\ensuremath{\Pmu^+\Pmu^-}}\xspace}

\def\taup       {{\ensuremath{\Ptau^+}}\xspace}
\def\taum       {{\ensuremath{\Ptau^-}}\xspace}

\def\ellell     {\ensuremath{\ell^+ \ell^-}\xspace}

\def\neu        {{\ensuremath{\Pnu}}\xspace}
\def\neub       {{\ensuremath{\overline{\Pnu}}}\xspace}

\def\g      {{\ensuremath{\Pgamma}}\xspace}

\def\W      {{\ensuremath{\PW}}\xspace}
\def\Wp     {{\ensuremath{\PW^+}}\xspace}
\def\Wm     {{\ensuremath{\PW^-}}\xspace}

\def\Z      {{\ensuremath{\PZ}}\xspace}

\def\quark     {{\ensuremath{\Pq}}\xspace}

\def\uquark    {{\ensuremath{\Pu}}\xspace}

\def\dquark    {{\ensuremath{\Pd}}\xspace}

\def\squark    {{\ensuremath{\Ps}}\xspace}
\def\squarkbar {{\ensuremath{\overline \squark}}\xspace}

\def\cquark    {{\ensuremath{\Pc}}\xspace}
\def\cquarkbar {{\ensuremath{\overline \cquark}}\xspace}

\def\bquark    {{\ensuremath{\Pb}}\xspace}
\def\bquarkbar {{\ensuremath{\overline \bquark}}\xspace}

\def\tquark    {{\ensuremath{\Pt}}\xspace}
\def\tquarkbar {{\ensuremath{\overline \tquark}}\xspace}

\def\pion   {{\ensuremath{\Ppi}}\xspace}

\def\pip    {{\ensuremath{\pion^+}}\xspace}
\def\pim    {{\ensuremath{\pion^-}}\xspace}

\def\kaon    {{\ensuremath{\PK}}\xspace}

\def\KorKbar {\kern \thebaroffset\optbar{\kern -\thebaroffset \PK}{}\xspace}
\def\Kz      {{\ensuremath{\kaon^0}}\xspace}

\def\Kp      {{\ensuremath{\kaon^+}}\xspace}
\def\Km      {{\ensuremath{\kaon^-}}\xspace}
\def\Kpm     {{\ensuremath{\kaon^\pm}}\xspace}

\def\KS      {{\ensuremath{\kaon^0_{\mathrm{S}}}}\xspace}

\def\Kstarz  {{\ensuremath{\kaon^{*0}}}\xspace}

\def\Kstar   {{\ensuremath{\kaon^*}}\xspace}

\def\Kstarp  {{\ensuremath{\kaon^{*+}}}\xspace}

\def\Dbar    {{\ensuremath{\offsetoverline{\PD}}}\xspace}
\def\D       {{\ensuremath{\PD}}\xspace}

\def\DorDbar {\kern \thebaroffset\optbar{\kern -\thebaroffset \PD}\xspace}

\def\Dzb     {{\ensuremath{\Dbar{}^0}}\xspace}
\def\Dp      {{\ensuremath{\D^+}}\xspace}
\def\Dm      {{\ensuremath{\D^-}}\xspace}

\def\DpDm    {\ensuremath{\Dp {\kern -0.16em \Dm}}\xspace}
\def\Dstar   {{\ensuremath{\D^*}}\xspace}

\def\Dstarp  {{\ensuremath{\D^{*+}}}\xspace}

\def\Ds      {{\ensuremath{\D^+_\squark}}\xspace}

\def\B       {{\ensuremath{\PB}}\xspace}

\def\BorBbar {\kern \thebaroffset\optbar{\kern -\thebaroffset \PB}\xspace}

\def\Bd      {{\ensuremath{\B^0}}\xspace}

\def\BdorBdbar {\kern \thebaroffset\optbar{\kern -\thebaroffset \Bd}\xspace}
\def\Bu      {{\ensuremath{\B^+}}\xspace}
\def\Bub     {{\ensuremath{\B^-}}\xspace}
\def\Bp      {{\ensuremath{\Bu}}\xspace}
\def\Bm      {{\ensuremath{\Bub}}\xspace}

\def\Bs      {{\ensuremath{\B^0_\squark}}\xspace}

\def\BsorBsbar {\kern \thebaroffset\optbar{\kern -\thebaroffset \Bs}\xspace}

\def\jpsi     {{\ensuremath{{\PJ\mskip -3mu/\mskip -2mu\Ppsi}}}\xspace}
\def\psitwos  {{\ensuremath{\Ppsi{(2S)}}}\xspace}

\def\Y#1S{\ensuremath{\PUpsilon{(#1S)}}\xspace}

\def\FourS {{\Y4S}\xspace}

\def\proton      {{\ensuremath{\Pp}}\xspace}

\def\Lz          {{\ensuremath{\PLambda}}\xspace}
\def\Lbar        {{\ensuremath{\offsetoverline{\PLambda}}}\xspace}
\def\LorLbar     {\kern \thebaroffset\optbar{\kern -\thebaroffset \PLambda}\xspace}

\def\Xires       {{\ensuremath{\PXi}}\xspace}

\def\Xiresbar       {{\ensuremath{\offsetoverline{\Xires}}}\xspace}

\def\Lb           {{\ensuremath{\Lz^0_\bquark}}\xspace}

\newcommand{\decay}[2]{\ensuremath{\mathinner{#1\!\to #2}}\xspace}

\def\to                 {\ensuremath{\rightarrow}\xspace}

\def\qsq       {{\ensuremath{q^2}}\xspace}

\def\CP                {{\ensuremath{C\!P}}\xspace}

\def\Vts  {{\ensuremath{V_{\tquark\squark}^{\phantom{\ast}}}}\xspace}
\def\Vub  {{\ensuremath{V_{\uquark\bquark}^{\phantom{\ast}}}}\xspace}
\def\Vcb  {{\ensuremath{V_{\cquark\bquark}^{\phantom{\ast}}}}\xspace}

\def\AT#1     {\ensuremath{A_{\mathrm{T}}^{#1}}\xspace}           

\def\Bsmm     {\decay{\Bs}{\mup\mun}}

\def\C#1      {\ensuremath{\mathcal{C}_{#1}}\xspace}                       
\def\Cp#1     {\ensuremath{\mathcal{C}_{#1}^{'}}\xspace}                    
\def\Ceff#1   {\ensuremath{\mathcal{C}_{#1}^{\mathrm{(eff)}}}\xspace}        
\def\Cpeff#1  {\ensuremath{\mathcal{C}_{#1}^{'\mathrm{(eff)}}}\xspace}       
\def\Ope#1    {\ensuremath{\mathcal{O}_{#1}}\xspace}                       
\def\Opep#1   {\ensuremath{\mathcal{O}_{#1}^{'}}\xspace}                    

\newcommand{\nospaceunit}[1]{\ensuremath{\text{#1}}}       
\newcommand{\aunit}[1]{\ensuremath{\text{\,#1}}}       

\newcommand{\tev}{\aunit{Te\kern -0.1em V}\xspace}
\newcommand{\gev}{\aunit{Ge\kern -0.1em V}\xspace}
\newcommand{\mev}{\aunit{Me\kern -0.1em V}\xspace}
\newcommand{\kev}{\aunit{ke\kern -0.1em V}\xspace}
\newcommand{\ev}{\aunit{e\kern -0.1em V}\xspace}
 
\newcommand{\mevc}{\ensuremath{\aunit{Me\kern -0.1em V\!/}c}\xspace}
\newcommand{\gevc}{\ensuremath{\aunit{Ge\kern -0.1em V\!/}c}\xspace}
\newcommand{\mevcc}{\ensuremath{\aunit{Me\kern -0.1em V\!/}c^2}\xspace}
\newcommand{\gevcc}{\ensuremath{\aunit{Ge\kern -0.1em V\!/}c^2}\xspace}

\def\mum  {\ensuremath{\,\upmu\nospaceunit{m}}\xspace}

\def\gsim{{~\raise.15em\hbox{$>$}\kern-.85em
          \lower.35em\hbox{$\sim$}~}\xspace}
\def\lsim{{~\raise.15em\hbox{$<$}\kern-.85em
          \lower.35em\hbox{$\sim$}~}\xspace}

\def\tell1  {TELL1\xspace}
\def\ukl1   {UKL1\xspace}

\newcommand{\lhcborcid}[1]{\href{https://orcid.org/#1}{\hspace*{0.1em}\raisebox{-0.45ex}{\includegraphics[width=1em]{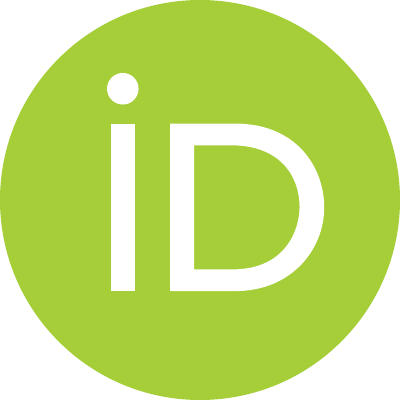}}}}

\hypersetup{
  colorlinks   = true, 
  urlcolor     = blue, 
  linkcolor    = blue, 
  citecolor    = red   
}

\ifEnableSectionTOCLinks
    \usepackage[explicit]{titlesec} 
    
    \let\oldcontentsline\contentsline
    \renewcommand

    \titleformat{\section}{\normalfont\Large\bf}{\hyperlink{tocsection.\thesection}{{\thesection} \parbox[t]{\dimexpr\textwidth-1pc}{#1}}}{1pc}{}

    \titleformat{\subsection}{\normalfont\bf}{\hyperlink{tocsubsection.\thesubsection}{{\thesubsection} \parbox[t]{\dimexpr\textwidth-1pc}{#1}}}{1pc}{}

    \titleformat{name=\section,numberless}[display]{}{}{0pt}{\normalfont\Huge\bfseries #1}
\fi

\usepackage{cite} 
\usepackage{LHCb/mciteplus}

\usepackage{longtable} 

\begin{document}

\renewcommand{\thefootnote}{\fnsymbol{footnote}}
\setcounter{footnote}{1}


\begin{titlepage}

\vspace*{-1.5cm}

\noindent
\begin{tabular*}{\linewidth}{lc@{\extracolsep{\fill}}r@{\extracolsep{0pt}}}
 & & Nikhef 2026-006  \\  
 & & \today \\ 
 & & \\
\hline
\end{tabular*}

\vspace*{4.0cm}

{\normalfont\bfseries\boldmath\huge
\begin{center}
  \papertitle
\end{center}
}

\vspace*{2.0cm}

\begin{center}
\paperauthors$^1$.
\bigskip\\
{\normalfont\itshape\footnotesize
$ ^1$Nikhef, Amsterdam, Netherlands\\
}
\end{center}

\vspace{\fill}

\begin{abstract}
  \noindent The next 20 years will be the golden age of flavour physics, with the operation of the
  LHCb and Belle~II experiments. After that an \epem collider could further improve the precision
  with sizeable \Z, \Wp\Wm and \tquark\tquarkbar runs.
\end{abstract}
\vspace*{2.0cm}
\begin{center}
  Proceedings of the International Workshop on Future Linear Colliders, LCWS'25\\ 20--24~Oct~2025, Valencia, Spain.
\end{center}
\vspace{\fill}
{\footnotesize
\centerline{\copyright~\papercopyright. \href{\paperlicenceurl}{\paperlicence}.}}
\vspace*{2mm}

\end{titlepage}

\pagestyle{empty}  


\newpage
\setcounter{page}{2}
\mbox{~}


\renewcommand{\thefootnote}{\arabic{footnote}}
\setcounter{footnote}{0}


\cleardoublepage


\pagestyle{plain} 
\setcounter{page}{1}
\pagenumbering{arabic}

\linenumbers


\section{Introduction}\label{sec1}
``Flavour physics highlights the virtue of virtuality'' wrote Eric Laenen
in the ESPPU Briefing Book~\cite{deBlas:2025gyz}. While colliders
access new physics by creating new particles, flavour and electroweak physics
provide tests that are sensitive to the ripples of new physics at much lower energies.

For instance the absence of \decay{\bquark\squarkbar}{\Z} couplings in the SM makes
flavour-changing neutral current processes sensitive to new physics that can hide in loops.
The \BmmKs process is a benchmark process that gives access to multiple angular distributions.
One of those, dubbed $P_5'$, is well known for deviating from the SM.
\begin{figure}[b]
\centering
\includegraphics[height=0.26\textheight]{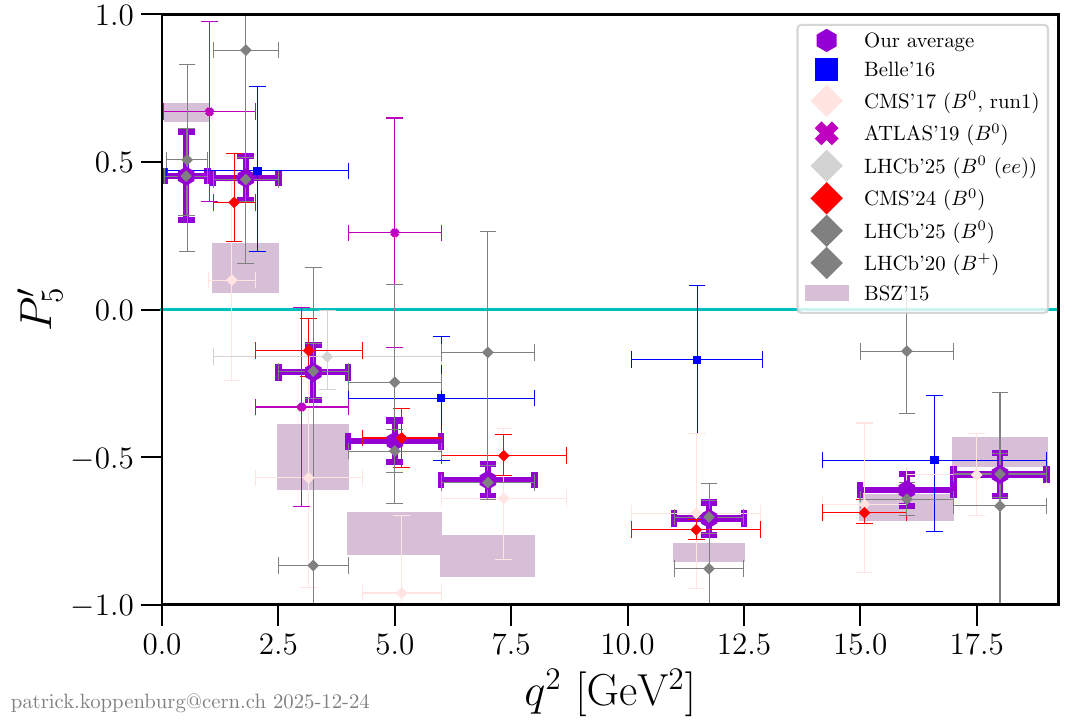} %
\includegraphics[height=0.26\textheight]{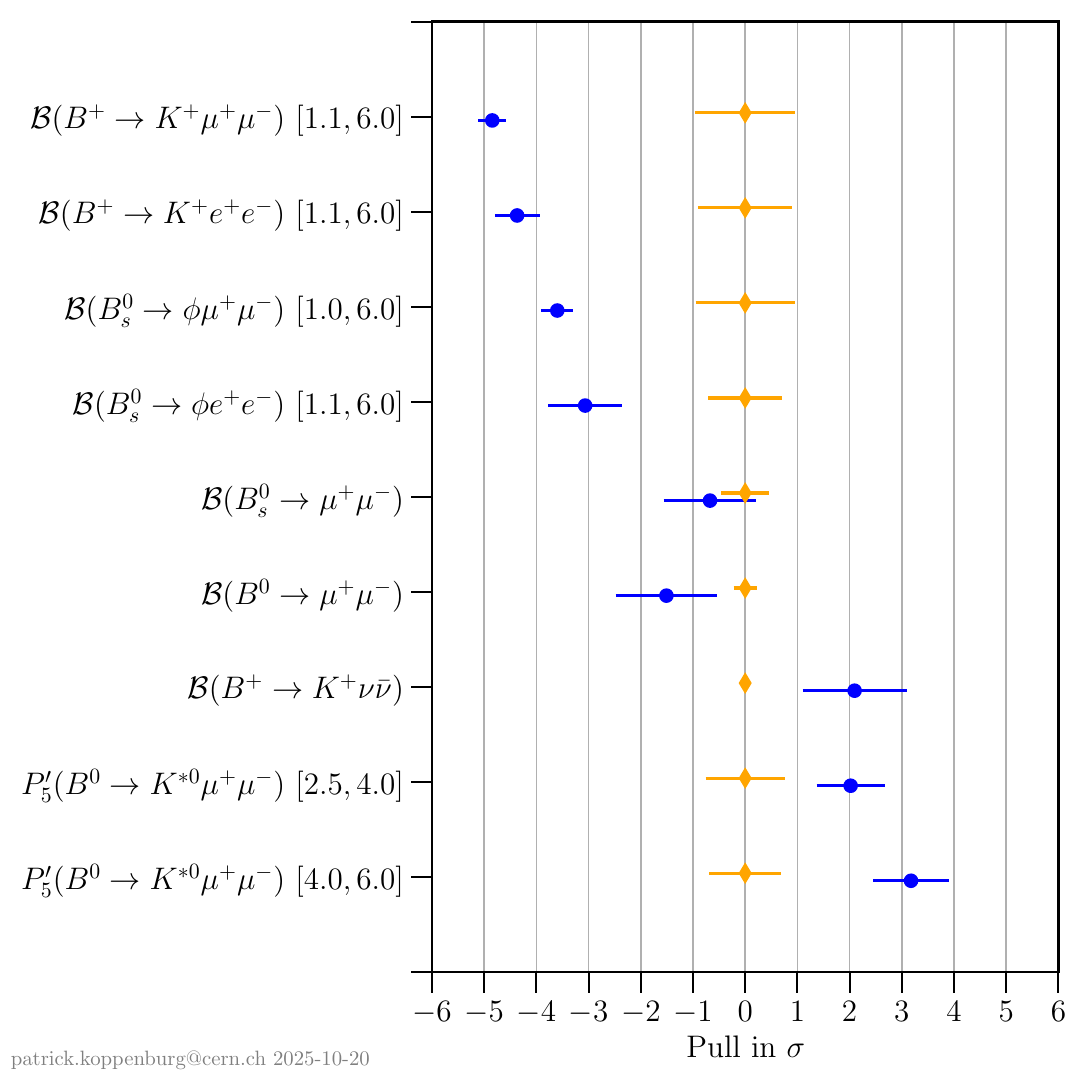} %
\caption{(left) Experimental results~\cite{LHCb-PAPER-2025-041,LHCb-PAPER-2024-022,CMS:2024atz,Belle:2016fev,CMS:2017rzx,ATLAS:2018gqc} for $P_5'$ in \BmmKs processes and their average compared to theory prediction~\cite{Alguero:2023jeh}. (right) Selected experimental results normalised and compared to the theory expectation~\cite{PKanomalies}.}\label{fig:Anomalies}\label{fig:BllKs}
\end{figure}
Figure~\ref{fig:BllKs}~(left) shows the most precise
measurements~\cite{LHCb-PAPER-2025-041,LHCb-PAPER-2024-022,CMS:2024atz,Belle:2016fev,CMS:2017rzx,ATLAS:2018gqc} versus the dilepton mass squared.
Also, multiple branching fractions are in tension with the Standard Model, as shown in Fig.~\ref{fig:Anomalies}~(right).
It is too early to say whether these effects are due to new physics of nonfactorisable QCD contributions,
but much larger samples will tell us.

Flavour physics also provides the only known source of \CP violation in the standard
model\footnote{Neutrinos will follow for sure.} via the KM mechanism~\cite{KM}. The unitarity conditions of this matrix
provide precision tests of the standard model as shown in Fig.~\ref{fig:CKM}.
There are also intriguing results in semileptonic \decay{b}{c\tau\neub} processes,
and a long-standing discrepancy in the determinations of the CKM matrix elements
\Vub and \Vcb by inclusive and exclusive methods~\cite{HFLAV23}.

\section{Future prospects at HL--LHC and Belle II}\label{sec:future}
\begin{figure}[t]
\centering
\includegraphics[height=0.19\textheight]{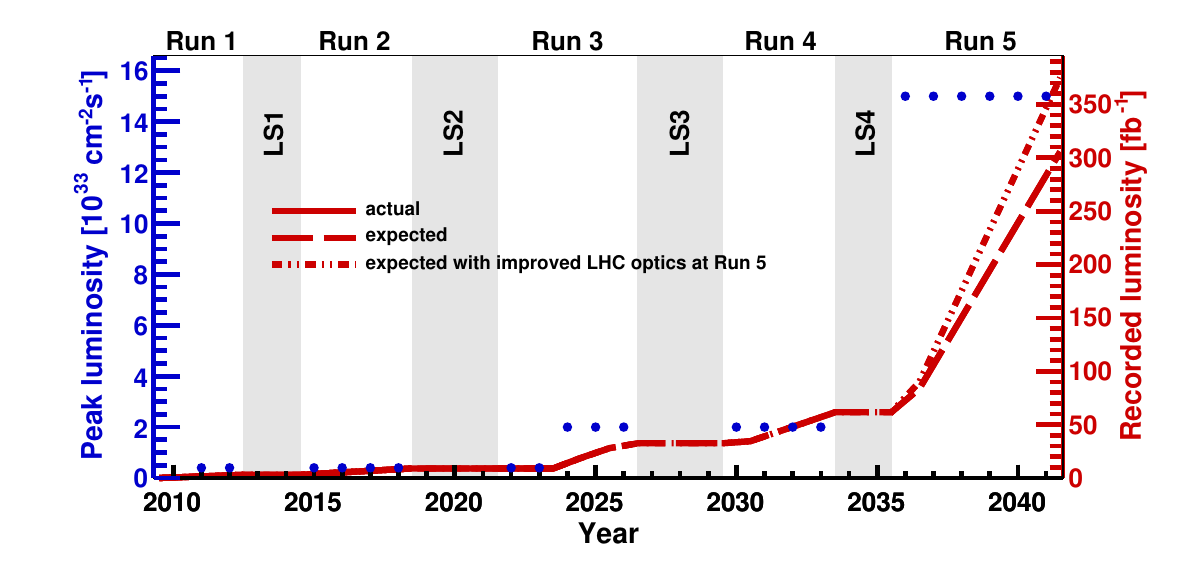} %
\includegraphics[height=0.19\textheight]{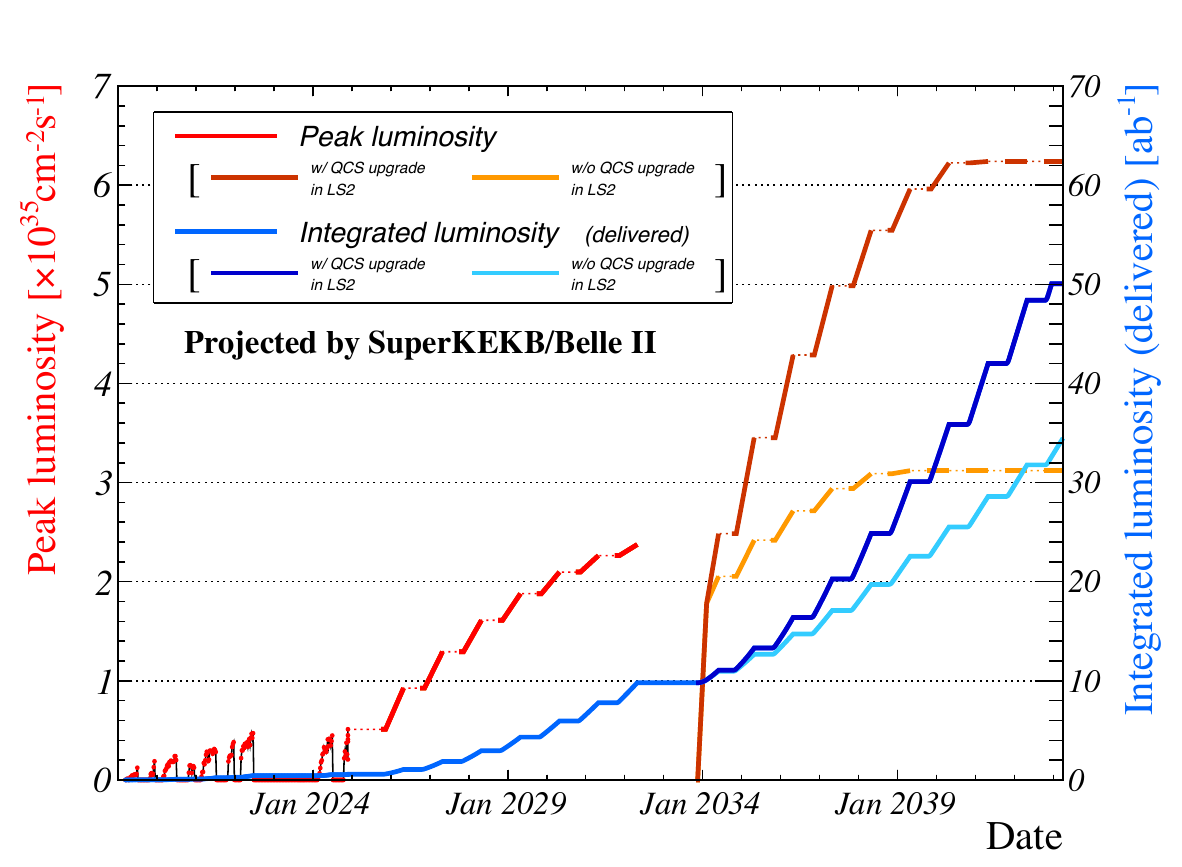}
\caption{Planned instantaneous and integrated luminosity (left) at LHCb~\cite{LHCb-TDR-023,LHCb-PII-Physics} and (right) Belle~II~\cite{belleIIlumi}.}\label{fig:LHCb}\label{fig:BelleII}
\end{figure}
Today flavour physics with \bquark- and \cquark-hadrons is dominated by the
LHCb experiment at the LHC and the Belle~II experiment at KEK-B. Both
recently had a substantial upgrade and are planning to collect much
larger samples that available now. The integrated luminosities are shown
in Fig.~\ref{fig:LHCb}. LHCb plans a second upgrade during the long shutdown 4.
The ``legacy'' LHCb sample collected up to 2018 corresponds to about 3\% of the target total integrated luminosity,
while Belle~II is planning to collect a sample 50 to 60 times larger than what its predecessor Belle
recorded. ATLAS and CMS will also continue to contribute to \bquark physics and profit from the large
luminosities at HL-LHC. The prospects are described in detail in Ref.~\cite{ATLAS:2025lrr}.

With such datasets the \decay{b}{s} Wilson coefficients can be determined
in small bins of the \BdmmKs dilepton mass \qsq.
If these coefficients come out \qsq-dependent, it would be an indication that QCD effects are at play.
On the other hand if physics is universal and different from the standard model, it would be a strong
sign of new physics.

\begin{figure}[b]%
\centering%
\includegraphics[width=0.5\textwidth]{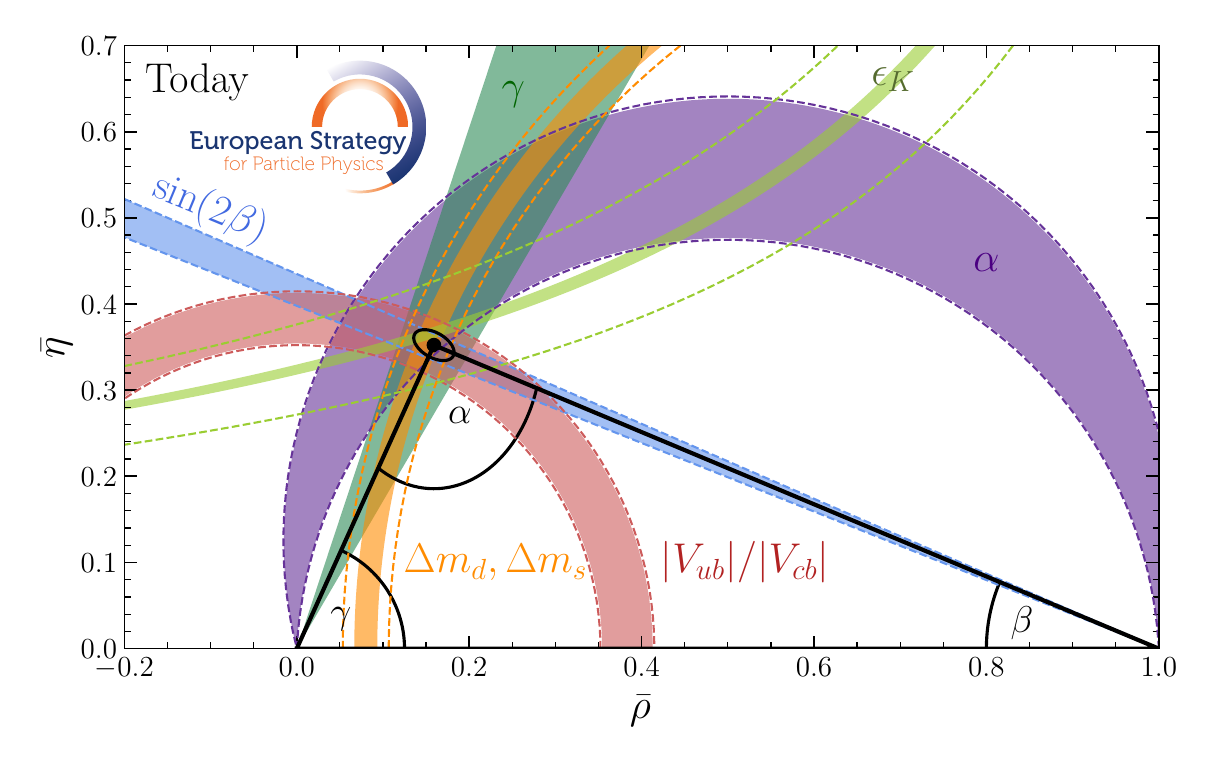}%
\includegraphics[width=0.5\textwidth]{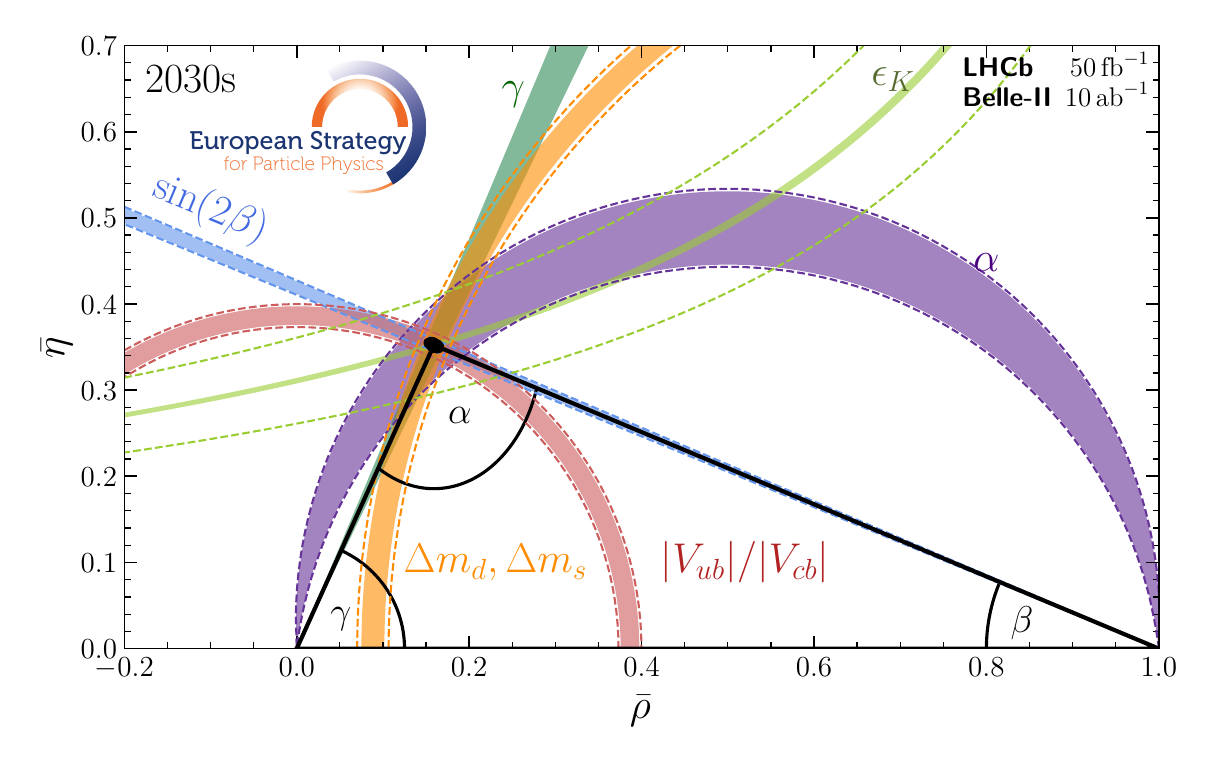}\\%
\includegraphics[width=0.5\textwidth]{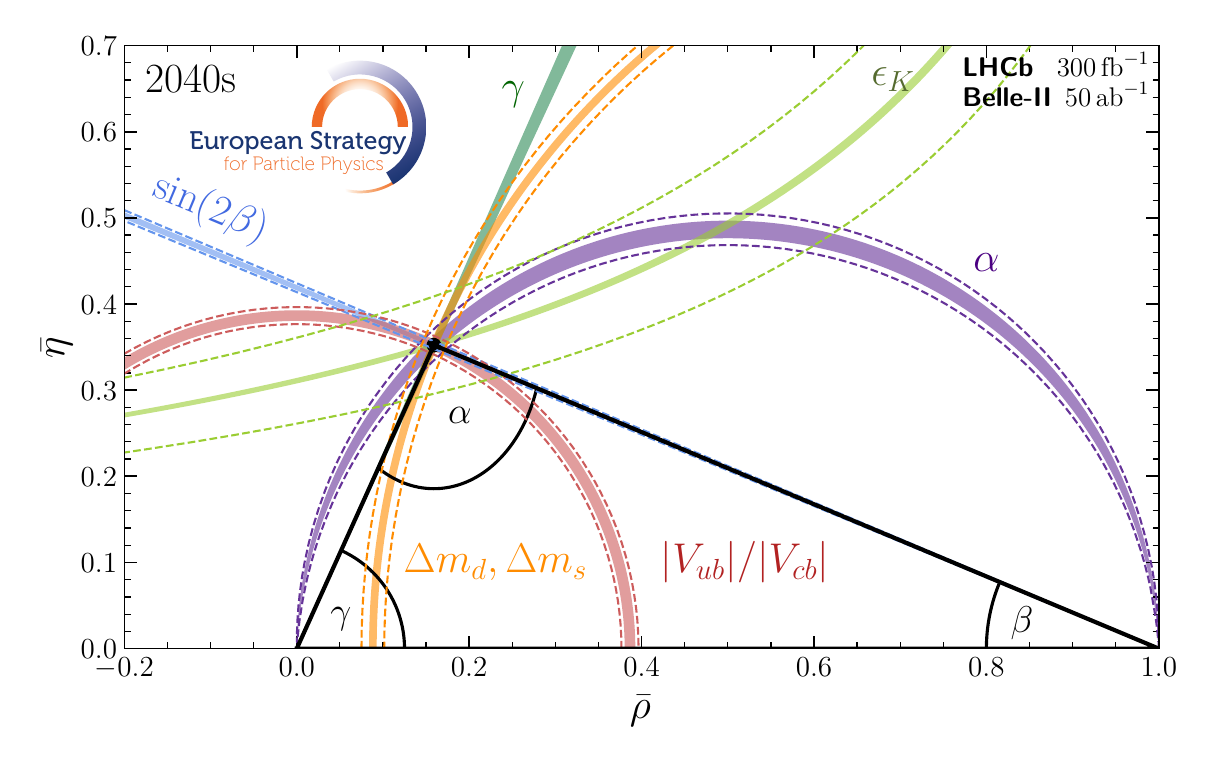}%
\includegraphics[width=0.5\textwidth]{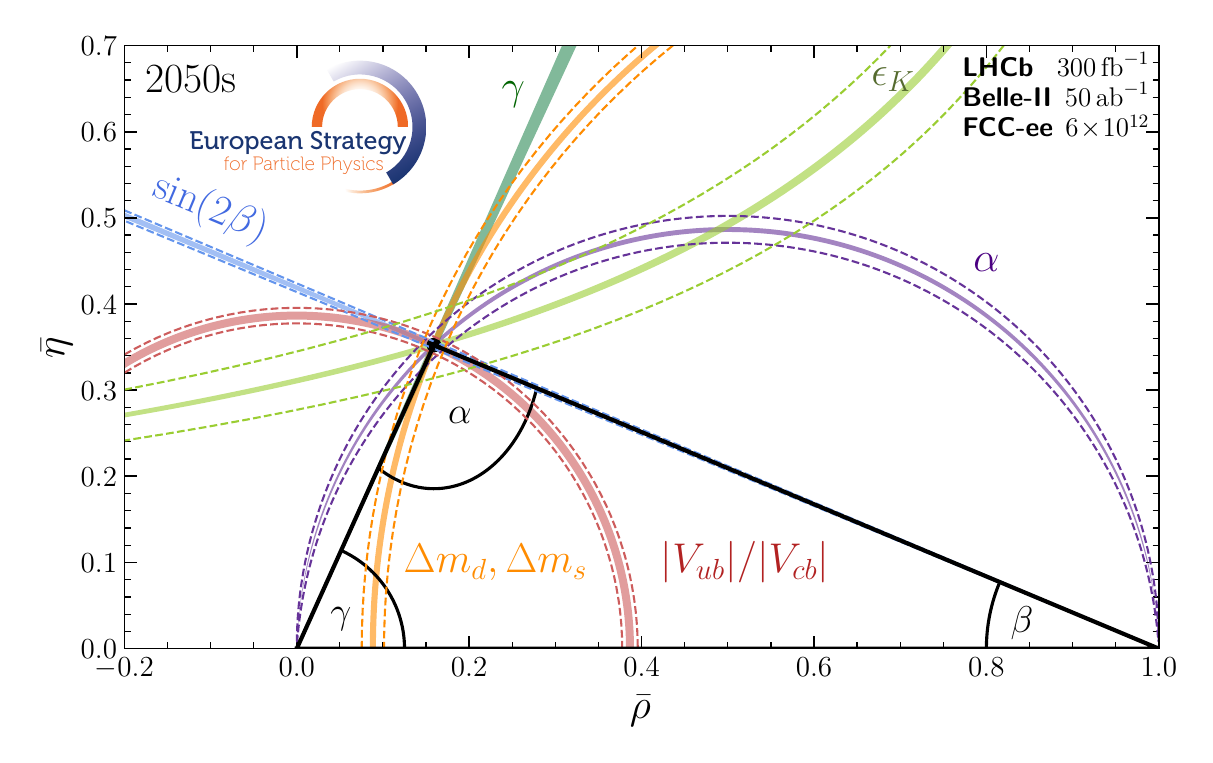}%
\caption{CKM unitarity conditions (top left) today, (top right) in the 2030s, %
(bottom left) after HL-LHC and Belle II and (bottom right) after FCC-ee. Plots from Ref.~\cite{deBlas:2025gyz}.}\label{fig:CKM}%
\end{figure}%
\begin{figure}[t]%
\centering%
\includegraphics[height=0.25\textheight]{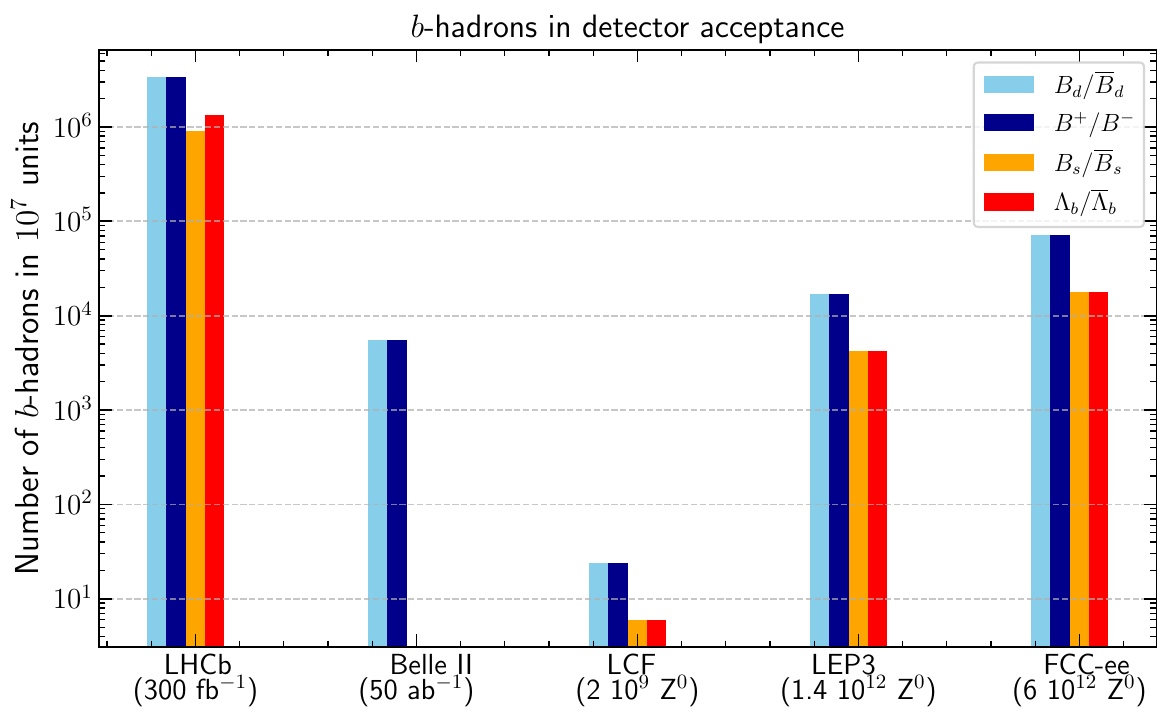} %
\includegraphics[height=0.25\textheight]{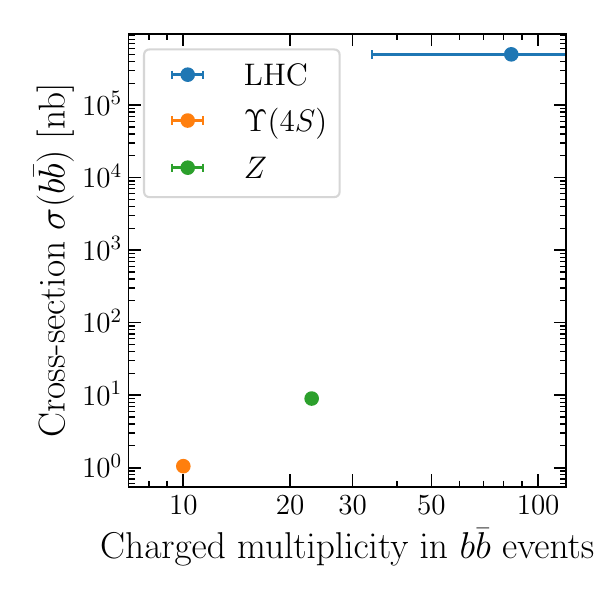} %
\caption{ (left) Average charged multiplicity in \bquark\bquarkbar events %
  at the \FourS and \Z resonances, and at the LHC.%
  }\label{fig:Mult}\label{fig:Cross}%
\end{figure}%
Similarly the precision on the CKM unitarity triangle will be greatly improved (Fig.~\ref{fig:CKM}).
\section{Flavour physics at the \Z pole}\label{sec:Z}
Owing to the lower cross-section, \bquark-physics at the \Z pole cannot compete
with the LHC in sheer numbers, see Fig.~\ref{fig:Cross}. However, the \Z environment
is cleaner, which permits a much improved tagging efficiency~\cite{Abidi:2025dfw}.
It is however less clean than runs at the \FourS where only the two \B mesons are produced
(Fig.~\ref{fig:Mult}). A \Z run is therefore somewhere in between LHC and Belle~II
both in terms of signal yields and background levels.

A tera-\Z run would exploit this environment, as well as very precise detectors, for
few selected processes as time-dependent \Bsmm or \decay{\Bs}{\phi\mumu},
\decay{b}{s\neu\neub}, \decay{\B}{\D^{(*)}\tau\neub}, 
as well as rare \Ptau decays~\cite{Abidi:2025dfw,deBlas:2025gyz}.

\begin{figure}[b]
\centering
\includegraphics[height=0.23\textheight]{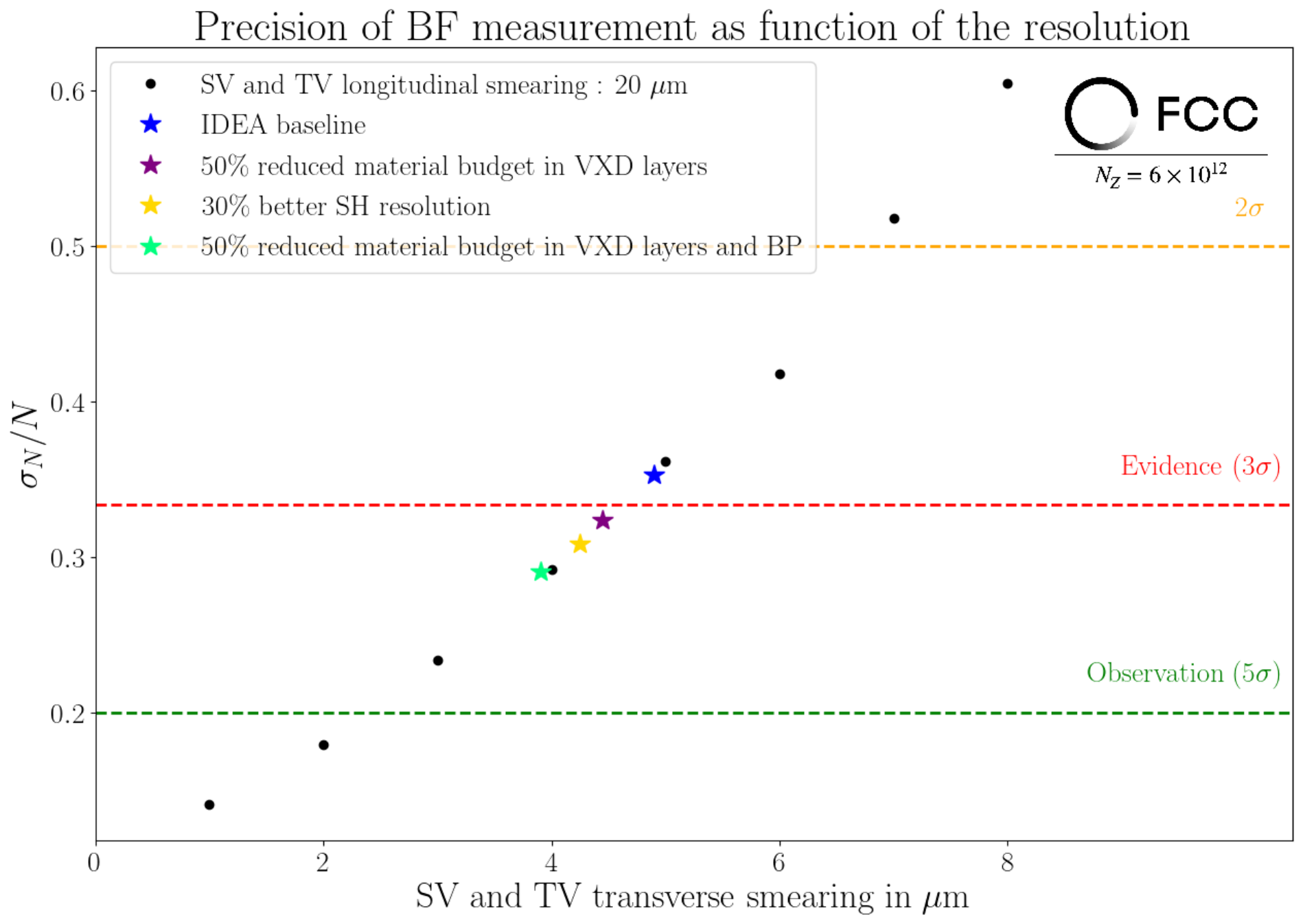} %
\includegraphics[height=0.23\textheight]{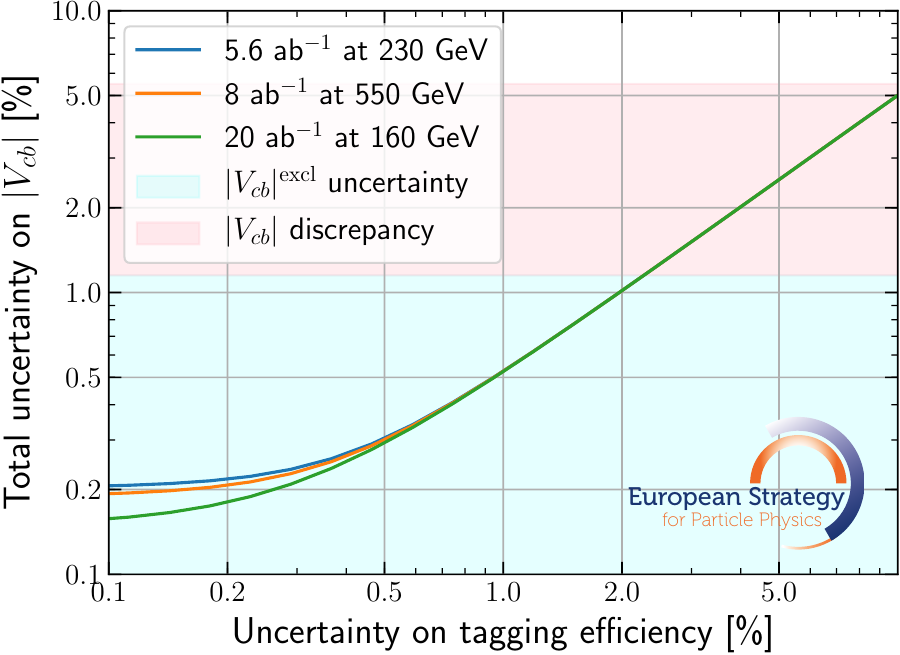}
\caption{(left) Precision on \decay{\Bd}{\Kstarz\taup\taum} versus multiple scattering
  effects at FCC~\cite{Abidi:2025dfw}. (right) Total precision on $|\Vcb|$
versus uncertainty on jet flavour-tagging in \W decays~\cite{deBlas:2025gyz}.}\label{fig:Vcb}\label{fig:Ktautau}
\end{figure}

A benchmark process is \decay{\Bd}{\Kstarz\taup\taum}, which is sensitive to
the same physics as \decay{\B}{\D^{(*)}\tau\neub} but at the loop level.
Assuming the SM, a run with $6\times10^{12}$ \Z bosons would just about allow for an evidence.
The situation improves with a very thin beam pipe restricting multiple scattering to
less than 2\mum, see Fig~\ref{fig:Ktautau}, and assuming all detectors are equally
equipped for flavour physics.

\section{Flavour physics beyond the \Z pole}\label{sec:W}
A close inspection of Fig.~\ref{fig:CKM} shows minor improvements of the CKM unitarity triangle
from a \Z run. The remaining dominating uncertainties
affect the sides of the triangle facing $\beta$, and the $\epsilon_K$ band. In both cases
the uncertainty relates to $|\Vub|$ and $|\Vcb|$ which are affected by the inclusive versus exclusive
discrepancy. A measurement of $|\Vcb|$ from \W decays rather than from semileptonic \bquark decays
would likely solve the
conundrum. Here however, the dominating uncertainly is of systematic nature,
related to the knowledge of the jet flavour-tagging efficiency, see Fig.~\ref{fig:Vcb}.

Similarly, the precision on the other triangle side is dominated by $|\Vts|$, which is obtained
from \Bs oscillations. A direct measurement could be obtained from \decay{t}{s\Wp} processes,
provided a significant run beyond the \tquark\tquarkbar threshold is
recorded.\footnote{The plot in Fig.~\ref{fig:CKM} is therefore mislabelled in
``$6\times10^{12}$'' as \W and \tquark runs are equally important.}

\section{Conclusion}\label{sec:Concl}
The golden age of flavour is now. With the Belle~II and LHCb Upgrade~II experiments
unprecedented precision will be obtained. An \epem collider could after that
address some remaining issues. For a full understanding large \Wp\Wm and \tquark\tquarkbar
runs are as important as a tera-\Z run. Additionally, all experiments must be
equipped for flavour physics, including sufficient PID capabilities and exquisite
vertexing.


\addcontentsline{toc}{section}{References}
\bibliographystyle{LHCb/LHCb}
\bibliography{exp,theory,LHCb/standard,LHCb/LHCb-PAPER,LHCb/LHCb-CONF,LHCb/LHCb-DP,LHCb/LHCb-TDR,LHCb/LHCb-PUB}

\end{document}